\documentclass[twocolumn,pra,superscriptaddress]{revtex4-1}
\usepackage{hyperref}
\usepackage{graphicx}
\usepackage{dcolumn}
\usepackage{bm}
\usepackage{tikz}
\usepackage{amsmath}
\usepackage{endnotes}
\begin{document}

\title{All entangled states can generate certified randomness}

\author{Xing Chen}
\email{xingchenphy@gmail.com}
\affiliation{Institute of Physics, Beijing National Laboratory for
  Condensed Matter Physics, Chinese Academy of Sciences, Beijing
  100190, China}

\begin{abstract}
Random number has many applications, it plays an important role in quantum information processing. It's not difficult to generate true random numbers, the main difficulty is how to certify the random numbers generated by untrusted devices. In [Nature(London) 464, 1021 (2010)], the authors provided us a way to generate certified random number by Bell's theorem. In their scheme, we can use the nonlocal behavior of entangled states to generate certified randomness. But there are entangled states, which admit a local hidden variable model, could not be used in their scheme. We show in our paper that the nonlocal correlations in every entangled state can be used to generate certified randomness, and we use Werner states as an example to show how to quantify the output randomness.
\end{abstract}

\maketitle

\section{Introduction}

\subsection{Background}
Random number has many important applications nowadays, such as quantum key distribution and the test of Bell's theorem. It is believed that we can't generate true random numbers by classical processes. On the other hand, the random numbers generated by quantum physics are truly random because of the superposition of quantum states. There are three kinds of quantum random number generators(QRNG)\cite{Ma2016}, the first kind is practical QRNG, which is built on trusted devices; the second kind is self-testing QRNG, which can generate true random numbers without trusting the measurement devices; the last kind of QRNG is called semi-self-testing QRNG, which partially combines the advantages of the first two QRNGs.

It is not difficult to generate true random numbers by tools of quantum physics currently. In the first kind of QRNG, we can actually generate random numbers at a very satisfied speed\cite{Xu:12}\cite{Yuan}\cite{youqi}\cite{mara2017}. But these true random numbers are not easy to be certified, because the adversaries may use memory-stick attack\cite{acin2016}. The so called memory-stick attack is implemented in the following way: the adversaries may generate a very long true random numbers and store them in the devices, when the users use this devices to generate random numbers, the generated random numbers are exactly what the adversaries stored in the devices.

Nonlocality can be used to generate certified random numbers \cite{phd2009}\cite{pironio2010}. In \cite{phd2009}\cite{pironio2010} the authors connected Bell's theorem with randomness, the violation of Bell's inequality guarantees that the generated random numbers contain true randomness, and this randomness is measurement device independent. Especially in \cite{pironio2010}, a lower bound of the output randomness was derived by the nonlocality in entangled states. However, not all entangled states can violate Bell's inequality, and those who admit a local hidden variable model can not be used in their randomness certification system. In order to take advantage of all entangled states, we need a new certification scheme. Inspired by previous work \cite{phd2009}\cite{pironio2010}, with the method provided by \cite{bus2012}\cite{denis}\cite{bra}, we connect randomness with entanglement. In our paper, we can generate certified private randomness by any entangled state, the output randomness in our protocol is measurement device independent and it only needs some fresh randomness as input.

\subsection{Related work}
Similar randomness generation protocol was also mentioned in Chaturvedi and Banik's paper\cite{anu2014}, but there are some major differences between their paper and our paper.

In their paper, Chaturvedi and Banik only gave a very specific entangled state's output randomness. Also, in their paper, they claimed that the output randomness was safe even local operations and classical communication(LOCC) was allowed between two measurement devices. However, the untrusted measurement devices could share some extra entangled states which are not known by Alice and Bob, and we show in our paper that these extra entangled states have the potential to damage the output randomness without being detected.

\subsection{The result of our paper}
In our paper, we give a general lower bound for our randomness generation protocol. The same as\cite{pironio2010}, the classical communication between different measurement devices is forbidden in our protocol. Moreover, the lower bound derived in our paper takes the secret entangled states shared between untrusted measurement devices into consideration.

Generally, the certified randomness in our scheme is different from the randomness verified by Bell's theorem. This is because the random numbers in these two protocols are not generated in the same way. In Bell's scenario, the randomness is created by measuring entangled states with optimal positive operator valued measure(POVM)(in order to obtain the maximum Bell value $I$), and the guessing probability of the measurement results is increasing with the decreasing of $I$, only when $I>2$, the guessing probability is less than $1$, so this protocol needs the violation of Bell's inequality to guarantee the randomness in the measurement results. While in our protocol, the inputs are nonorthogonal states, the maximum guessing probability has an upper bound which is always less than $1$, and the measurement results in our protocol can be verified by any entangled state, so every entangled state is a useful resource in our protocol.

Our paper is organised in the following way. In section II we give a brief introduction to the randomness generation protocol mentioned in \cite{pironio2010}. In section III, the main part of our paper, we show how to construct the protocol in our paper, and we give an example to show how to quantify the output randomness. Section IV is a brief conclusion.

\section{Randomness certified by Bell's theorem}
\begin{figure}[hbt!]
  \centering
  \includegraphics[width=5cm]{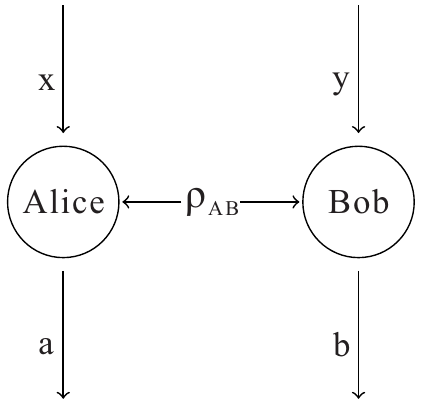}
\caption{Randomness certified by Bell nonlocality}\label{Bellrandom}
\end{figure}

Before the illustration of our randomness generation protocol, we first give a brief introduction to the scheme mentioned in \cite{pironio2010}, the authors used Clauser-Horne-Shimony-Holt (CHSH)\cite{bel}\cite{cla}correlation function
\begin{equation}
\label{chsh}
  I=\sum_{x,y}(-1)^{xy}[P(a=b|xy)-P(a\neq b|xy)]
\end{equation}
as an example to show how to generate certified randomness. Where $x$ and $y$ represent certain type of measurement, such as the different polarization directions of polarizers. $a$ and $b$ are the measurement results of Alice and Bob. Choosing certain $x$ and $y$ to measure the entangled state between Alice and Bob could make the Bell value $I>2$, which means the violation of CHSH inequality, and the violation of CHSH inequality guarantees the randomness in the measurement results $a,b$. The randomness generation structure is shown in FIG.\ref{Bellrandom}

In this randomness generation protocol we can use the random seed $x$ and $y$ to generate certified random strings $a$ and $b$ if their correlation violate CHSH inequality, or more generally Bell's inequality.(One thing we should note here is that the input $x$ and $y$ should both be true random numbers, if they can be predicted by some adversaries, the adversaries may pre-create some correlated random numbers and store them into Alice and Bob's measurement devices to implement memory-stick attack.) The private randomness in the output strings $a$ and $b$ is quantified by min-entropy, and the details could be find in the supplementary information of \cite{pironio2010}.

\section{Randomness certified by any entangled state}

This section is the main part of our paper. It shows how to generate certified randomness by any entangled state, and how to quantify the output randomness.

The randomness generation protocol mentioned in section II takes advantage of the Bell nonlocality in entangled states. However, some entangled states may admit a local hidden variable model, and they do not contain Bell nonlocality\cite{wer}\cite{bru}. In order to make every entangled state useful in certified randomness generation protocol, we must use other property of entangled states instead of Bell nonlocality. In \cite{bus2012}, the author proved that with nonorthogonal states as input, the measurement correlation in every entangled state cannot be created by any separate state with local operations and shared randomness(LOSR). The protocol introduced in our paper uses this property of entangled state. For simplicity, we use bipartite scenario to illustrate our protocol.

First, we introduce a Bell-like inequality\cite{denis}. In bipartite scenario, Alice and Bob are two separated parties, they share an entangled state $\rho_{AB}$, then they both input some quantum states $\{\tau_s\}$ and $\{\omega_t\}$ into the measurement devices, the measurement devices then output measurement results $a$ and $b$. The correlation between $a$ and $b$ is represented as $P(a,b|\tau_s,\omega_t)$.
\begin{equation}
  \label{correlation}
  P(a,b|\tau_s,\omega_t)=Tr[(P_a^{A'A}\otimes Q_b^{B'B})(\tau_s^{A'}\otimes\rho_{AB}\otimes\omega_t^{B'})]
\end{equation}
where $A,B$ is the Hilbert space of state $\rho_{AB}$, $A',B'$ is the Hilbert space of nonorthogonal states $\tau_s,\omega_t$, and $P_a^{A'A}, Q_b^{B'B}$ are the POVMs chosen by Alice and Bob. By making suitable joint measurement on their respective part of $\rho_{AB}$ and on the input quantum states $\tau_s,\omega_t$ with $P_a^{A'A}, Q_b^{B'B}$, Alice and Bob can obtain the correlation $P(a,b|\tau_s,\omega_t)$, which cannot be explained without entanglement\cite{bus2012}\cite{bra}. This allows the existence of the following linear combination of $P(a,b|\tau_s,\omega_t)$ :
\begin{equation}
\label{Belllike}
  I_{\rho_{AB}}=\sum_{a,b,s,t}\beta_{s,t,a,b}P(a,b|\tau_s,\omega_t)<0
\end{equation}
where $\beta_{s,t,a,b}$ are some real coefficients. The coefficients $\beta_{s,t,a,b}$ could be obtained from the decomposition of entanglement witness. Because of the completeness of $\{\tau_s\}$ and $\{\omega_t\}$, the entanglement witness $W$ of $\rho_{AB}$ could be decomposed as
\begin{equation}
\label{decompose}
  W=\sum_{s,t}\beta_{s,t}\tau_s^{\top}\otimes\omega_t^{\top}
\end{equation}
with $\beta_{s,t,1,1}=\beta_{s,t}$, and $\beta_{s,t,a,b}=0, (a,b)\neq(1,1)$, formula \ref{Belllike} becomes:
\begin{equation}
\label{Bellew}
  I_{\rho_{AB}}=\sum_{s,t}\beta_{s,t}P(1,1|\tau_s,\omega_t)<0
\end{equation}
It is proved in \cite{bra} that this inequality is greater than or equal to zero for any separated state with any possible POVM. For any entangled state, with the following POVMs,
\begin{equation}
\label{povm}
\begin{split}
  P_1^{A'A}=|\Phi_{d_A}^+\rangle\langle\Phi_{d_A}^+|,P_0^{A'A}=\textbf{I}-P_1^{A'A}\\
  Q_1^{B'B}=|\Phi_{d_B}^+\rangle\langle\Phi_{d_B}^+|,Q_0^{B'B}=\textbf{I}-Q_1^{B'B}
\end{split}
\end{equation}
where $|\Phi_{d_A}^+\rangle=\sum_{i=0}^{d_A-1}|i\rangle\otimes|i\rangle/\sqrt{d_A}$, and $|\Phi_{d_B}^+\rangle=\sum_{j=0}^{d_B-1}|j\rangle\otimes|j\rangle/\sqrt{d_B}$. The Bell-like value $I_{\rho_{AB}}$ will be:
\begin{equation}
\begin{split}
 I_{\rho_{AB}}=&\sum_{s,t}\beta_{s,t}P(1,1|\tau_s,\omega_t)\\
 =&\frac{Tr(W \rho_{AB})}{d_A d_B}\\
 <&0
\end{split}
\end{equation}

which is less than zero for entangled states, this means the measurement results $a,b$ must be unpredictable, otherwise there would be contradiction in quantum theory\cite{bra}.

\begin{figure}[hbt!]
  \centering
  \includegraphics[width=5cm]{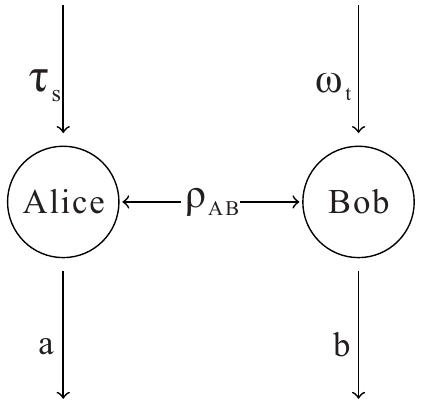}
\caption{Randomness certified by any entangled state. In this protocol, the input $\tau_s,\omega_t$ are nonorthogonal states, and the measurement devices of Alice and Bob are not allowed to communicate with each other. As long as the Bell-like value $I_{\rho_{AB}}$ is less than zero, Alice and Bob could extract certified randomness from the measurement results $a$ and $b$.}\label{Belllikerandom}
\end{figure}

With the above Bell-like correlation \ref{Bellew}, we can design a certified randomness generation protocol, where any entangled state is useful to generate certified randomness. The protocol is shown in FIG.\ref{Belllikerandom}
.
In this protocol, Alice and Bob randomly input some nonorthogonal states $\tau_s,\omega_t$ to the measurement devices, and they measure these nonorthogonal states jointly with their respective part of $\rho_{AB}$ by the POVM \ref{povm}. If the linear combination $I_{\rho_{AB}}$ of the correlation $P(1,1|\tau_s,\omega_t)$ is less than zero, then the output strings $a,b$ contain randomness which could not be predicted by the untrusted measurement devices. The certified randomness in $a,b$ is quantified by min-entropy $H_{\infty}(AB|ST)$.

We take Werner states as an example to show how to get the lower bound of $H_{\infty}(AB|ST)$. For Werner states,
\begin{equation}
\label{werner}
  \rho_z = \frac{1-z}{4} \textbf{I} + z|\Phi^+\rangle\langle\Phi^+|
\end{equation}
where $|\Phi^+\rangle$ is Bell state $(|00\rangle+|11\rangle)/\sqrt{2}$. The entanglement witness is:
\begin{equation}
  \label{witness}
  W_{\rho_z}=\frac12\textbf{I}-|\Phi^+\rangle\langle\Phi^+|
\end{equation}

The input quantum states $\tau_s$ and $\omega_t$ are nonorthogonal, they could be any nonorthogonal state which forms a complete basis. Without losing generality, they could be:
\begin{equation}
\label{input}
\begin{split}
 \tau_s=(\textbf{I}+\vec{v_s}\cdot\vec{\sigma})/2\\
 \omega_{t}=(\textbf{I}+\vec{v_t}\cdot\vec{\sigma})/2,
\end{split}
\end{equation}
where $\vec{\sigma}=(\sigma_x,\sigma_y,\sigma_z)$, $s,t\in\{0,1,2,3\}$, $\vec{v_0}=(1,1,1)/\sqrt{3},\vec{v_1}=(1,-1,-1)/\sqrt{3},\vec{v_2}=(-1,1,-1)/\sqrt{3},\vec{v_3}=(-1,-1,1)/\sqrt{3}$.

Next, we discuss the guessing probability of the measurement results. It is illustrated in the Appendix that the average guessing probability of the measurement results $a,b$ can be written as:
\begin{equation}
\label{pguess}
   p_{guess}=\frac{I_{\rho'}-I_{\rho_{z}}}{I_{\rho'}}\times 1+\frac{I_{\rho_{z}}}{I_{\rho'}}\times maxP_{\rho'}(a,b|\tau_s,\omega_t)
\end{equation}
And the upper bound of this average guessing probability $p_{guess}$ is
\begin{equation}\nonumber
\label{upper}
   p_{guess}\leq\frac{I_{\rho_{optimal}}-I_{\rho_{z}}}{I_{\rho_{optimal}}}+\frac{I_{\rho_{z}}}{I_{\rho_{optimal}}}\times P_{\rho_{max}}(a,b|\tau_s,\omega_t)
\end{equation}

For Werner states $\rho_z$, $I_{\rho_z}=(1-3z)/16$, $I_{\rho_{optimal}}=I_{|\Phi^+\rangle}= -1/8$. $P_{\rho_{max}}(a,b|\tau_s,\omega_t)$ is the maximum guessing probability with input \ref{input} and POVM \ref{povm}, it is proved in the Appendix that $P_{\rho_{max}}(a,b|\tau_s,\omega_t)\leq\frac{9+\sqrt{3}}{16}$. Then the upper bound of the average guessing probability is
\begin{equation}
\begin{split}
   p_{guess}\leq &\frac{I_{\rho_{|\Phi^+\rangle}}-I_{\rho_{z}}}{I_{\rho_{|\Phi^+\rangle}}}+\frac{I_{\rho_{z}}}{I_{\rho_{|\Phi^+\rangle}}}\times \frac{9+\sqrt{3}}{16}\\
   =&\frac{-1/8-(1-3z)/16}{-1/8}+\frac{(1-3z)/16}{-1/8}\times \frac{9+\sqrt{3}}{16}\\
   =&1-\frac{7-\sqrt{3}}{16}\times\frac{3z-1}{2}
\end{split}
\end{equation}

\begin{figure}[t!]
  \centering
  \includegraphics[width=\linewidth]{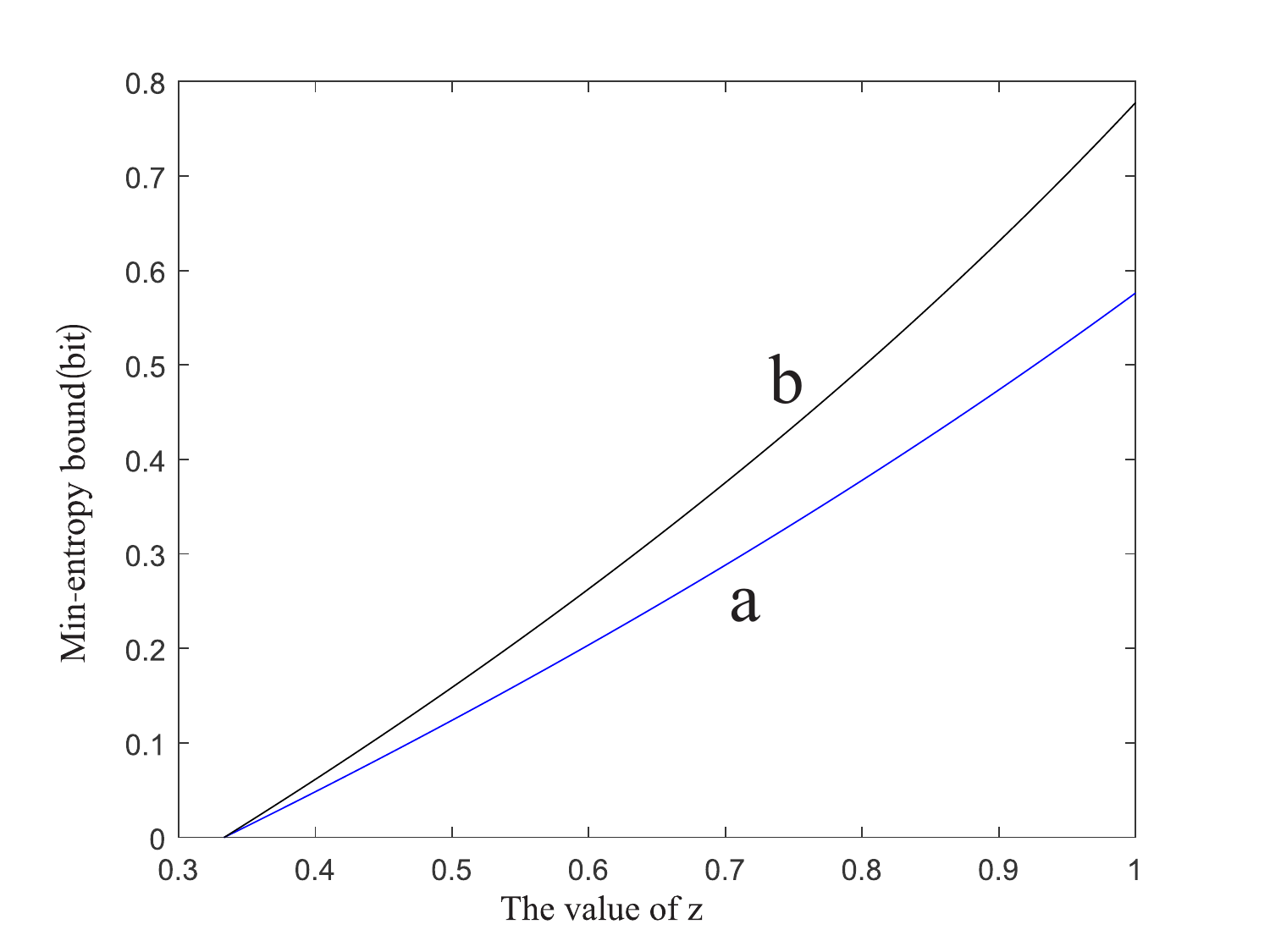}
  \caption{The lower bound of output randomness for Werner states $\rho_z$. The analytical upper bound of the average guessing probability is $p_{guess}\leq1-\frac{7-\sqrt{3}}{16}\times\frac{3z-1}{2}$, and this average guessing probability is the optimal one allowed by quantum theory. Then we can generate at least $-log_2(1-\frac{7-\sqrt{3}}{16}\times\frac{3z-1}{2})$ bits of certified randomness, and this randomness is the curve $a$ in the figure. Curve $b$ represents the lower bound obtained from semidefinite programs(SDP).}\label{output}
\end{figure}

According to the definition of min-entropy, the lower bound of the output randomness for Werner states $\rho_z$ is
\begin{equation}
  H_{\infty}(AB|ST)\geq -log_2(1-\frac{7-\sqrt{3}}{16}\times\frac{3z-1}{2})
\end{equation}
This lower bound is shown in FIG.\ref{output}.

From FIG.\ref{output} we can see that, for any entangled Werner state, the output randomness is greater than zero, which means any entanglement is useful in our randomness generation protocol.

\begin{figure}[t]
  \centering
  \includegraphics[width=\linewidth]{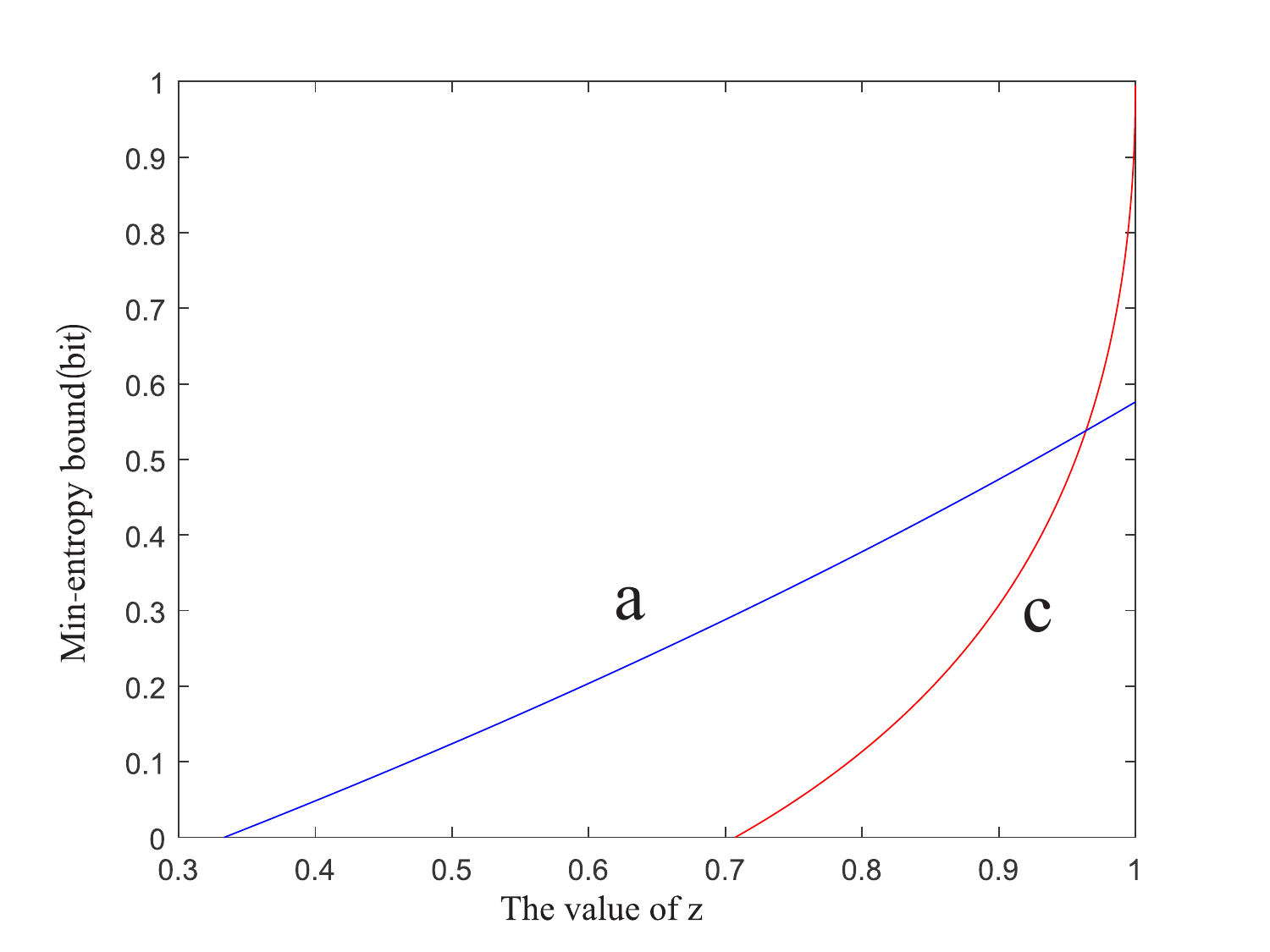}
  \caption{The compare of output certified randomness of $\rho_z$ in our protocol and in Bell protocol\cite{pironio2010}. The lower bound of our protocol is given by $ H_{\infty}(AB|ST)\geq -log_2(1-\frac{7-\sqrt{3}}{16}\times\frac{3z-1}{2})$, and is represented as blue line(curve $a$) in the figure. The red line(curve $c$) is the lower bound of output randomness certified by Bell's theorem, and its formula is $H_{\infty}(AB|XY)\geq[1-log_2(1+\sqrt{2-I^2/4})]$, the relationship between $z$ and $I$ is $I=2\sqrt{2}z$.}\label{compare}
\end{figure}

Comparing the lower bound of our randomness generation protocol with the analytical lower bound of the output randomness in \cite{pironio2010}, our randomness generation protocol could take advantage of all entangled states by introducing nonorthogonal states as input, and in low degree entanglement, our protocol could generate more secured randomness. The compare of the output randomness is shown in FIG.\ref{compare}

There are some constrains in our randomness generation protocol. Firstly, the corresponding entanglement witness must be decomposable by the nonorthogonal states and the nonorthogonal states cannot be distinguished by the measurement devices. These conditions require that the input quantum states $\tau_s$ and $\omega_t$ must be nonorthogonal states which form a complete basis, and the labels of these nonorthogonal states must not be revealed to the measurement devices. Secondly, it is more reasonable to trust the source devices other than the measurement devices\cite{bra}, so the generating devices of the nonorthogonal states should be trusted. Thirdly, any classical communication between the measurement devices is forbidden.

\section{Conclusion}
In this article, we construct a protocol to generate certified randomness by any entangled state. Similar protocol was also presented in \cite{anu2014}, but the authors didn't give a general lower bound of their protocol, and the lower bound obtained in\cite{anu2014} is not convincing without considering the secret entanglement shared between untrusted measurement devices. In our article, we take this situation into consideration and we obtain a general lower bound for this kind of randomness generation protocol.

Comparing to the protocol in \cite{pironio2010}, our protocol can take advantage of any entangled state, and our protocol is based on entanglement theory, the correctness of our protocol is guaranteed by the validity of quantum physics. The randomness generation protocol in our paper shows us the deep connection between entanglement and randomness, and further open questions may arise, such as what's the relationship between randomness and entanglement, are they equal to each other? How do nonorthogonal states assist entangled states, which admit a local hidden variable model, to generate certified randomness?
\bibliography{My-Collection}

\begin{thebibliography}{21}%
\makeatletter
\providecommand \@ifxundefined [1]{%
 \@ifx{#1\undefined}
}%
\providecommand \@ifnum [1]{%
 \ifnum #1\expandafter \@firstoftwo
 \else \expandafter \@secondoftwo
 \fi
}%
\providecommand \@ifx [1]{%
 \ifx #1\expandafter \@firstoftwo
 \else \expandafter \@secondoftwo
 \fi
}%
\providecommand \natexlab [1]{#1}%
\providecommand \enquote  [1]{``#1''}%
\providecommand \bibnamefont  [1]{#1}%
\providecommand \bibfnamefont [1]{#1}%
\providecommand \citenamefont [1]{#1}%
\providecommand \href@noop [0]{\@secondoftwo}%
\providecommand \href [0]{\begingroup \@sanitize@url \@href}%
\providecommand \@href[1]{\@@startlink{#1}\@@href}%
\providecommand \@@href[1]{\endgroup#1\@@endlink}%
\providecommand \@sanitize@url [0]{\catcode `\\12\catcode `\$12\catcode
  `\&12\catcode `\#12\catcode `\^12\catcode `\_12\catcode `\%12\relax}%
\providecommand \@@startlink[1]{}%
\providecommand \@@endlink[0]{}%
\providecommand \url  [0]{\begingroup\@sanitize@url \@url }%
\providecommand \@url [1]{\endgroup\@href {#1}{\urlprefix }}%
\providecommand \urlprefix  [0]{URL }%
\providecommand \Eprint [0]{\href }%
\providecommand \doibase [0]{http://dx.doi.org/}%
\providecommand \selectlanguage [0]{\@gobble}%
\providecommand \bibinfo  [0]{\@secondoftwo}%
\providecommand \bibfield  [0]{\@secondoftwo}%
\providecommand \translation [1]{[#1]}%
\providecommand \BibitemOpen [0]{}%
\providecommand \bibitemStop [0]{}%
\providecommand \bibitemNoStop [0]{.\EOS\space}%
\providecommand \EOS [0]{\spacefactor3000\relax}%
\providecommand \BibitemShut  [1]{\csname bibitem#1\endcsname}%
\let\auto@bib@innerbib\@empty
\bibitem [{\citenamefont {Ma}\ \emph {et~al.}(2016)\citenamefont {Ma},
  \citenamefont {Yuan}, \citenamefont {Cao}, \citenamefont {Qi},\ and\
  \citenamefont {Zhang}}]{Ma2016}%
  \BibitemOpen
  \bibfield  {author} {\bibinfo {author} {\bibfnamefont {X.}~\bibnamefont
  {Ma}}, \bibinfo {author} {\bibfnamefont {X.}~\bibnamefont {Yuan}}, \bibinfo
  {author} {\bibfnamefont {Z.}~\bibnamefont {Cao}}, \bibinfo {author}
  {\bibfnamefont {B.}~\bibnamefont {Qi}}, \ and\ \bibinfo {author}
  {\bibfnamefont {Z.}~\bibnamefont {Zhang}},\ }\href {\doibase
  10.1038/npjqi.2016.21} {\bibfield  {journal} {\bibinfo  {journal} {Nature
  Publishing Group}\ ,\ \bibinfo {pages} {1}} (\bibinfo {year}
  {2016})}\BibitemShut {NoStop}%
\bibitem [{\citenamefont {Xu}\ \emph {et~al.}(2012)\citenamefont {Xu},
  \citenamefont {Qi}, \citenamefont {Ma}, \citenamefont {Xu}, \citenamefont
  {Zheng},\ and\ \citenamefont {Lo}}]{Xu:12}%
  \BibitemOpen
  \bibfield  {author} {\bibinfo {author} {\bibfnamefont {F.}~\bibnamefont
  {Xu}}, \bibinfo {author} {\bibfnamefont {B.}~\bibnamefont {Qi}}, \bibinfo
  {author} {\bibfnamefont {X.}~\bibnamefont {Ma}}, \bibinfo {author}
  {\bibfnamefont {H.}~\bibnamefont {Xu}}, \bibinfo {author} {\bibfnamefont
  {H.}~\bibnamefont {Zheng}}, \ and\ \bibinfo {author} {\bibfnamefont {H.-K.}\
  \bibnamefont {Lo}},\ }\href {\doibase 10.1364/OE.20.012366} {\bibfield
  {journal} {\bibinfo  {journal} {Opt. Express}\ }\textbf {\bibinfo {volume}
  {20}},\ \bibinfo {pages} {12366} (\bibinfo {year} {2012})}\BibitemShut
  {NoStop}%
\bibitem [{\citenamefont {Yuan}\ \emph {et~al.}(2014)\citenamefont {Yuan},
  \citenamefont {Lucamarini}, \citenamefont {Dynes}, \citenamefont {Fröhlich},
  \citenamefont {Plews},\ and\ \citenamefont {Shields}}]{Yuan}%
  \BibitemOpen
  \bibfield  {author} {\bibinfo {author} {\bibfnamefont {Z.~L.}\ \bibnamefont
  {Yuan}}, \bibinfo {author} {\bibfnamefont {M.}~\bibnamefont {Lucamarini}},
  \bibinfo {author} {\bibfnamefont {J.~F.}\ \bibnamefont {Dynes}}, \bibinfo
  {author} {\bibfnamefont {B.}~\bibnamefont {Fröhlich}}, \bibinfo {author}
  {\bibfnamefont {A.}~\bibnamefont {Plews}}, \ and\ \bibinfo {author}
  {\bibfnamefont {A.~J.}\ \bibnamefont {Shields}},\ }\href {\doibase
  10.1063/1.4886761} {\bibfield  {journal} {\bibinfo  {journal} {Applied
  Physics Letters}\ }\textbf {\bibinfo {volume} {104}},\ \bibinfo {pages}
  {261112} (\bibinfo {year} {2014})},\ \Eprint
  {http://arxiv.org/abs/http://dx.doi.org/10.1063/1.4886761}
  {http://dx.doi.org/10.1063/1.4886761} \BibitemShut {NoStop}%
\bibitem [{\citenamefont {Nie}\ \emph {et~al.}(2015)\citenamefont {Nie},
  \citenamefont {Huang}, \citenamefont {Liu}, \citenamefont {Payne},
  \citenamefont {Zhang},\ and\ \citenamefont {Pan}}]{youqi}%
  \BibitemOpen
  \bibfield  {author} {\bibinfo {author} {\bibfnamefont {Y.-Q.}\ \bibnamefont
  {Nie}}, \bibinfo {author} {\bibfnamefont {L.}~\bibnamefont {Huang}}, \bibinfo
  {author} {\bibfnamefont {Y.}~\bibnamefont {Liu}}, \bibinfo {author}
  {\bibfnamefont {F.}~\bibnamefont {Payne}}, \bibinfo {author} {\bibfnamefont
  {J.}~\bibnamefont {Zhang}}, \ and\ \bibinfo {author} {\bibfnamefont {J.-W.}\
  \bibnamefont {Pan}},\ }\href {\doibase 10.1063/1.4922417} {\bibfield
  {journal} {\bibinfo  {journal} {Review of Scientific Instruments}\ }\textbf
  {\bibinfo {volume} {86}},\ \bibinfo {pages} {063105} (\bibinfo {year}
  {2015})},\ \Eprint {http://arxiv.org/abs/http://dx.doi.org/10.1063/1.4922417}
  {http://dx.doi.org/10.1063/1.4922417} \BibitemShut {NoStop}%
\bibitem [{\citenamefont {Marangon}\ \emph {et~al.}(2017)\citenamefont
  {Marangon}, \citenamefont {Vallone},\ and\ \citenamefont
  {Villoresi}}]{mara2017}%
  \BibitemOpen
  \bibfield  {author} {\bibinfo {author} {\bibfnamefont {D.~G.}\ \bibnamefont
  {Marangon}}, \bibinfo {author} {\bibfnamefont {G.}~\bibnamefont {Vallone}}, \
  and\ \bibinfo {author} {\bibfnamefont {P.}~\bibnamefont {Villoresi}},\ }\href
  {\doibase 10.1103/PhysRevLett.118.060503} {\bibfield  {journal} {\bibinfo
  {journal} {Phys. Rev. Lett.}\ }\textbf {\bibinfo {volume} {118}},\ \bibinfo
  {pages} {060503} (\bibinfo {year} {2017})}\BibitemShut {NoStop}%
\bibitem [{\citenamefont {Ac{\'\i}n}\ and\ \citenamefont
  {Masanes}(2016)}]{acin2016}%
  \BibitemOpen
  \bibfield  {author} {\bibinfo {author} {\bibfnamefont {A.}~\bibnamefont
  {Ac{\'\i}n}}\ and\ \bibinfo {author} {\bibfnamefont {L.}~\bibnamefont
  {Masanes}},\ }\href@noop {} {\bibfield  {journal} {\bibinfo  {journal}
  {Nature}\ }\textbf {\bibinfo {volume} {540}},\ \bibinfo {pages} {213}
  (\bibinfo {year} {2016})}\BibitemShut {NoStop}%
\bibitem [{\citenamefont {Colbeck}(2009)}]{phd2009}%
  \BibitemOpen
  \bibfield  {author} {\bibinfo {author} {\bibfnamefont {R.}~\bibnamefont
  {Colbeck}},\ }\href@noop {} {\bibfield  {journal} {\bibinfo  {journal} {arXiv
  preprint arXiv:0911.3814}\ } (\bibinfo {year} {2009})}\BibitemShut {NoStop}%
\bibitem [{\citenamefont {Pironio}\ \emph {et~al.}(2010)\citenamefont
  {Pironio}, \citenamefont {Ac{\'\i}n}, \citenamefont {Massar}, \citenamefont
  {de~La~Giroday}, \citenamefont {Matsukevich}, \citenamefont {Maunz},
  \citenamefont {Olmschenk}, \citenamefont {Hayes}, \citenamefont {Luo},
  \citenamefont {Manning} \emph {et~al.}}]{pironio2010}%
  \BibitemOpen
  \bibfield  {author} {\bibinfo {author} {\bibfnamefont {S.}~\bibnamefont
  {Pironio}}, \bibinfo {author} {\bibfnamefont {A.}~\bibnamefont {Ac{\'\i}n}},
  \bibinfo {author} {\bibfnamefont {S.}~\bibnamefont {Massar}}, \bibinfo
  {author} {\bibfnamefont {A.~B.}\ \bibnamefont {de~La~Giroday}}, \bibinfo
  {author} {\bibfnamefont {D.~N.}\ \bibnamefont {Matsukevich}}, \bibinfo
  {author} {\bibfnamefont {P.}~\bibnamefont {Maunz}}, \bibinfo {author}
  {\bibfnamefont {S.}~\bibnamefont {Olmschenk}}, \bibinfo {author}
  {\bibfnamefont {D.}~\bibnamefont {Hayes}}, \bibinfo {author} {\bibfnamefont
  {L.}~\bibnamefont {Luo}}, \bibinfo {author} {\bibfnamefont {T.~A.}\
  \bibnamefont {Manning}},  \emph {et~al.},\ }\href@noop {} {\bibfield
  {journal} {\bibinfo  {journal} {Nature}\ }\textbf {\bibinfo {volume} {464}},\
  \bibinfo {pages} {1021} (\bibinfo {year} {2010})}\BibitemShut {NoStop}%
\bibitem [{\citenamefont {Buscemi}(2012)}]{bus2012}%
  \BibitemOpen
  \bibfield  {author} {\bibinfo {author} {\bibfnamefont {F.}~\bibnamefont
  {Buscemi}},\ }\href {\doibase 10.1103/PhysRevLett.108.200401} {\bibfield
  {journal} {\bibinfo  {journal} {Phys. Rev. Lett.}\ }\textbf {\bibinfo
  {volume} {108}},\ \bibinfo {pages} {200401} (\bibinfo {year}
  {2012})}\BibitemShut {NoStop}%
\bibitem [{\citenamefont {Rosset}\ \emph {et~al.}(2013)\citenamefont {Rosset},
  \citenamefont {Branciard}, \citenamefont {Gisin},\ and\ \citenamefont
  {Liang}}]{denis}%
  \BibitemOpen
  \bibfield  {author} {\bibinfo {author} {\bibfnamefont {D.}~\bibnamefont
  {Rosset}}, \bibinfo {author} {\bibfnamefont {C.}~\bibnamefont {Branciard}},
  \bibinfo {author} {\bibfnamefont {N.}~\bibnamefont {Gisin}}, \ and\ \bibinfo
  {author} {\bibfnamefont {Y.-C.}\ \bibnamefont {Liang}},\ }\href
  {http://stacks.iop.org/1367-2630/15/i=5/a=053025} {\bibfield  {journal}
  {\bibinfo  {journal} {New Journal of Physics}\ }\textbf {\bibinfo {volume}
  {15}},\ \bibinfo {pages} {053025} (\bibinfo {year} {2013})}\BibitemShut
  {NoStop}%
\bibitem [{\citenamefont {Branciard}\ \emph {et~al.}(2013)\citenamefont
  {Branciard}, \citenamefont {Rosset}, \citenamefont {Liang},\ and\
  \citenamefont {Gisin}}]{bra}%
  \BibitemOpen
  \bibfield  {author} {\bibinfo {author} {\bibfnamefont {C.}~\bibnamefont
  {Branciard}}, \bibinfo {author} {\bibfnamefont {D.}~\bibnamefont {Rosset}},
  \bibinfo {author} {\bibfnamefont {Y.-C.}\ \bibnamefont {Liang}}, \ and\
  \bibinfo {author} {\bibfnamefont {N.}~\bibnamefont {Gisin}},\ }\href
  {\doibase 10.1103/PhysRevLett.110.060405} {\bibfield  {journal} {\bibinfo
  {journal} {Phys. Rev. Lett.}\ }\textbf {\bibinfo {volume} {110}},\ \bibinfo
  {pages} {060405} (\bibinfo {year} {2013})}\BibitemShut {NoStop}%
\bibitem [{\citenamefont {Chaturvedi}\ and\ \citenamefont
  {Banik}(2015)}]{anu2014}%
  \BibitemOpen
  \bibfield  {author} {\bibinfo {author} {\bibfnamefont {A.}~\bibnamefont
  {Chaturvedi}}\ and\ \bibinfo {author} {\bibfnamefont {M.}~\bibnamefont
  {Banik}},\ }\href {http://stacks.iop.org/0295-5075/112/i=3/a=30003}
  {\bibfield  {journal} {\bibinfo  {journal} {EPL (Europhysics Letters)}\
  }\textbf {\bibinfo {volume} {112}},\ \bibinfo {pages} {30003} (\bibinfo
  {year} {2015})}\BibitemShut {NoStop}%
\bibitem [{\citenamefont {Bell}(1964)}]{bel}%
  \BibitemOpen
  \bibfield  {author} {\bibinfo {author} {\bibfnamefont {J.~S.}\ \bibnamefont
  {Bell}},\ }\href@noop {} {\enquote {\bibinfo {title} {On the einstein
  podolsky rosen paradox},}\ } (\bibinfo {year} {1964})\BibitemShut {NoStop}%
\bibitem [{\citenamefont {Clauser}\ \emph {et~al.}(1969)\citenamefont
  {Clauser}, \citenamefont {Horne}, \citenamefont {Shimony},\ and\
  \citenamefont {Holt}}]{cla}%
  \BibitemOpen
  \bibfield  {author} {\bibinfo {author} {\bibfnamefont {J.~F.}\ \bibnamefont
  {Clauser}}, \bibinfo {author} {\bibfnamefont {M.~A.}\ \bibnamefont {Horne}},
  \bibinfo {author} {\bibfnamefont {A.}~\bibnamefont {Shimony}}, \ and\
  \bibinfo {author} {\bibfnamefont {R.~A.}\ \bibnamefont {Holt}},\ }\href
  {\doibase 10.1103/PhysRevLett.23.880} {\bibfield  {journal} {\bibinfo
  {journal} {Phys. Rev. Lett.}\ }\textbf {\bibinfo {volume} {23}},\ \bibinfo
  {pages} {880} (\bibinfo {year} {1969})}\BibitemShut {NoStop}%
\bibitem [{\citenamefont {Werner}(1989)}]{wer}%
  \BibitemOpen
  \bibfield  {author} {\bibinfo {author} {\bibfnamefont {R.~F.}\ \bibnamefont
  {Werner}},\ }\href {\doibase 10.1103/PhysRevA.40.4277} {\bibfield  {journal}
  {\bibinfo  {journal} {Phys. Rev. A}\ }\textbf {\bibinfo {volume} {40}},\
  \bibinfo {pages} {4277} (\bibinfo {year} {1989})}\BibitemShut {NoStop}%
\bibitem [{\citenamefont {Brunner}\ \emph {et~al.}(2014)\citenamefont
  {Brunner}, \citenamefont {Cavalcanti}, \citenamefont {Pironio}, \citenamefont
  {Scarani},\ and\ \citenamefont {Wehner}}]{bru}%
  \BibitemOpen
  \bibfield  {author} {\bibinfo {author} {\bibfnamefont {N.}~\bibnamefont
  {Brunner}}, \bibinfo {author} {\bibfnamefont {D.}~\bibnamefont {Cavalcanti}},
  \bibinfo {author} {\bibfnamefont {S.}~\bibnamefont {Pironio}}, \bibinfo
  {author} {\bibfnamefont {V.}~\bibnamefont {Scarani}}, \ and\ \bibinfo
  {author} {\bibfnamefont {S.}~\bibnamefont {Wehner}},\ }\href {\doibase
  10.1103/RevModPhys.86.419} {\bibfield  {journal} {\bibinfo  {journal} {Rev.
  Mod. Phys.}\ }\textbf {\bibinfo {volume} {86}},\ \bibinfo {pages} {419}
  (\bibinfo {year} {2014})}\BibitemShut {NoStop}%
\bibitem [{\citenamefont {Chen}\ \emph {et~al.}(2017)\citenamefont {Chen},
  \citenamefont {Hu},\ and\ \citenamefont {Zhou}}]{chen2017}%
  \BibitemOpen
  \bibfield  {author} {\bibinfo {author} {\bibfnamefont {X.}~\bibnamefont
  {Chen}}, \bibinfo {author} {\bibfnamefont {X.}~\bibnamefont {Hu}}, \ and\
  \bibinfo {author} {\bibfnamefont {D.~L.}\ \bibnamefont {Zhou}},\ }\href
  {\doibase 10.1103/PhysRevA.95.052326} {\bibfield  {journal} {\bibinfo
  {journal} {Phys. Rev. A}\ }\textbf {\bibinfo {volume} {95}},\ \bibinfo
  {pages} {052326} (\bibinfo {year} {2017})}\BibitemShut {NoStop}%
\bibitem [{\citenamefont {Horn}(1962)}]{horn1962}%
  \BibitemOpen
  \bibfield  {author} {\bibinfo {author} {\bibfnamefont {A.}~\bibnamefont
  {Horn}},\ }\href {http://projecteuclid.org/euclid.pjm/1103036720} {\bibfield
  {journal} {\bibinfo  {journal} {Pacific J. Math.}\ }\textbf {\bibinfo
  {volume} {12}},\ \bibinfo {pages} {225} (\bibinfo {year} {1962})}\BibitemShut
  {NoStop}%
\bibitem [{\citenamefont {Vandenberghe}\ and\ \citenamefont
  {Boyd}(1996)}]{sdp1996}%
  \BibitemOpen
  \bibfield  {author} {\bibinfo {author} {\bibfnamefont {L.}~\bibnamefont
  {Vandenberghe}}\ and\ \bibinfo {author} {\bibfnamefont {S.}~\bibnamefont
  {Boyd}},\ }\href@noop {} {\bibfield  {journal} {\bibinfo  {journal} {SIAM
  review}\ }\textbf {\bibinfo {volume} {38}},\ \bibinfo {pages} {49} (\bibinfo
  {year} {1996})}\BibitemShut {NoStop}%
\bibitem [{\citenamefont {Navascu\'es}\ \emph {et~al.}(2007)\citenamefont
  {Navascu\'es}, \citenamefont {Pironio},\ and\ \citenamefont
  {Ac\'{\i}n}}]{sdp2007}%
  \BibitemOpen
  \bibfield  {author} {\bibinfo {author} {\bibfnamefont {M.}~\bibnamefont
  {Navascu\'es}}, \bibinfo {author} {\bibfnamefont {S.}~\bibnamefont
  {Pironio}}, \ and\ \bibinfo {author} {\bibfnamefont {A.}~\bibnamefont
  {Ac\'{\i}n}},\ }\href {\doibase 10.1103/PhysRevLett.98.010401} {\bibfield
  {journal} {\bibinfo  {journal} {Phys. Rev. Lett.}\ }\textbf {\bibinfo
  {volume} {98}},\ \bibinfo {pages} {010401} (\bibinfo {year}
  {2007})}\BibitemShut {NoStop}%
\bibitem [{\citenamefont {Lim}(2016)}]{c2016}%
  \BibitemOpen
  \bibfield  {author} {\bibinfo {author} {\bibfnamefont {C.~C.~W.}\
  \bibnamefont {Lim}},\ }\href {\doibase 10.1103/PhysRevA.93.020101} {\bibfield
   {journal} {\bibinfo  {journal} {Phys. Rev. A}\ }\textbf {\bibinfo {volume}
  {93}},\ \bibinfo {pages} {020101} (\bibinfo {year} {2016})}\BibitemShut
  {NoStop}%
\end{thebibliography}%
\bibliographystyle{apsrev4-1}
\newpage
\appendix
\section*{Appendix}
In this Appendix, we will show how to get the lower bound of the output randomness in our randomness generation protocol.
\section{The average guessing probability of measurement results}
\setcounter{section}{1}
The POVMs in our protocol are:
\begin{equation}
\label{povm2}
\begin{split}
  P_1^{A'A}=|\Phi_{d_A}^+\rangle\langle\Phi_{d_A}^+|,P_0^{A'A}=\textbf{I}-P_1^{A'A}\\
  Q_1^{B'B}=|\Phi_{d_B}^+\rangle\langle\Phi_{d_B}^+|,Q_0^{B'B}=\textbf{I}-Q_1^{B'B}
\end{split}
\end{equation}
and the input nonorthogonal states are
\begin{equation}
\label{noninput}
\begin{split}
 \tau_s=(\textbf{I}+\vec{v_s}\cdot\vec{\sigma})/2\\
 \omega_{t}=(\textbf{I}+\vec{v_t}\cdot\vec{\sigma})/2,
\end{split}
\end{equation}
where $\vec{\sigma}=(\sigma_x,\sigma_y,\sigma_z)$, $s,t=0,1,2,3$, $\vec{v_0}=(1,1,1)/\sqrt{3},\vec{v_1}=(1,-1,-1)/\sqrt{3},\vec{v_2}=(-1,1,-1)/\sqrt{3},\vec{v_3}=(-1,-1,1)/\sqrt{3}$.

The average guessing probability of the measurement results is :
\begin{equation}\nonumber
\label{pguess2}
\begin{split}
   p_{guess}=&\frac{I_{\rho'}-I_{\rho_{AB}}}{I_{\rho'}}\times 1+\frac{I_{\rho_{AB}}}{I_{\rho'}}\times maxP_{\rho'}(a,b|\tau_s,\omega_t)\\
   \leq & \frac{I_{\rho_{optimal}}-I_{\rho_{AB}}}{I_{\rho_{optimal}}}+\frac{I_{\rho_{AB}}}{I_{\rho_{optimal}}}\times maxP_{\rho'}(a,b|\tau_s,\omega_t)\\
   \leq & \frac{I_{\rho_{optimal}}-I_{\rho_{AB}}}{I_{\rho_{optimal}}}+\frac{I_{\rho_{AB}}}{I_{\rho_{optimal}}}\times P_{\rho_{max}}(a,b|\tau_s,\omega_t)
\end{split}
\end{equation}
where $I_{\rho'}$ is the Bell-like value of entangled state $\rho'$, which shares the same given entanglement witness $W$ of $\rho_{AB}$. $P_{\rho'}(a,b|\tau_s,\omega_t)$ is the correlation of the measurement results. We have $I_{\rho'}\leq I_{\rho_{AB}}<0$ and $maxP_{\rho'}(a,b|\tau_s,\omega_t)\geq maxP_{\rho_{AB}}(a,b|\tau_s,\omega_t)$, otherwise, it's meaningless for the measurement devices to forge the measurement results. $I_{\rho_{optimal}}$ is the minimum Bell-like value for entanglement witness $W$. $P_{\rho_{max}}(a,b|\tau_s,\omega_t))$ is the maximum measurement correlation for any two-qubit state with POVM (\ref{povm2}) and with nonorthogonal inputs (\ref{noninput})

Next, we give an example to show the meaning of the average guessing probability. Suppose the entangled state shared between Alice and Bob is
\begin{equation}
\label{werner2}
  \rho_{z}= \frac{1-z}{4} \textbf{I} + z|\Phi^+\rangle\langle\Phi^+|
\end{equation}
the corresponding entanglement witness will be
\begin{equation}
  \label{witness}
  W_{\rho_z}=\frac12\textbf{I}-|\Phi^+\rangle\langle\Phi^+|
\end{equation}
With the POVM (\ref{povm2}) and with nonorthogonal inputs (\ref{noninput}),the correlations of measurement results are
\begin{eqnarray}\nonumber
\begin{split}
 &P_{\rho_z}(a,b|\tau_s,\omega_t)=zP_{|\Phi^+\rangle}(a,b,|\tau_s,\omega_t)+(1-z) P_{\textbf{I}}(a,b|\tau_s,\omega_t)\\
& =z\frac{7-5a-5b+4ab}{12}+(1-z)\frac{(3-2a)(3-2b)}{16}\\
&=\frac{(3-2a)(3-2b)}{16}+z\frac{(1-2a)(1-2b)}{48}
\end{split}
\end{eqnarray}
and the maximum $P_{\rho_z}(a,b|\tau_s,\omega_t)$ is
\begin{equation}
\label{guess}
\begin{split}
 &max_{a,b}P_{\rho_z}(a,b|\tau_s,\omega_t)\\
 &=P_{\rho_z}(0,0|\tau_s,\omega_t)\\
&=\frac{9}{16}+z\frac{1}{48}\\
&=\frac{27+z}{48}
\end{split}
\end{equation}
This equation shows that the guessing probability is becoming larger with the increasing of $z$. It seems without entanglement, the output randomness is better. Actually, this is not the case. From formula (\ref{guess}), the guessing probability is maximum when the shared entangled state is Bell state, so the optimal cheating strategy for the measurement devices is to fake the measurement results with built-in state $|\Phi^+\rangle$. Here is the detail, suppose $z=0.34$, then
\begin{equation}
\label{share2}
  \rho_{z_0}=0.66\times\frac{\textbf{I}}{4}  + 0.34\times|\Phi^+\rangle\langle\Phi^+|
\end{equation}
If the measurement devices do the corresponding measurement honestly, the maximum guessing probability of the measurement results will be
\begin{equation}
\label{guess2}
\begin{split}
 &max_{a,b}P_{\rho_{z_0}}(a,b|\tau_s,\omega_t)\\
 &=P_{\rho_{z_0}}(0,0|\tau_s,\omega_t)\\
 &=\frac{27+0.34}{48}
\end{split}
\end{equation}
the output randomness is $-log_2(\frac{27+0.34}{48})=0.812$, it's quite large for a low degree entangled state like (\ref{share2}). However, most part of the output randomness cannot be certified, the untrusted measurement devices could use the following trick to jeopardize the security in the measurement results:

When Alice and Bob used entangled state (\ref{share2}) to generate certified random number, the expected Bell-like value $I_{\rho_{z_0}}$ is
\begin{equation}
\label{werdetail}
\begin{split}
  I_{\rho_{z_0}}&=\frac{Tr(W\rho_{z_0})}{4}\\
  &=\frac{1-3\times0.34}{16}\\
  &=-\frac1{800}
\end{split}
\end{equation}

Without being detected, the measurement devices could take advantage of their built-in Bell state $|\Phi^+\rangle$ to deceive Alice and Bob. The entanglement witness of $|\Phi^+\rangle$ and $\rho_{z_0}$ could both be $W_{\rho_z}$, the corresponding Bell-like value of $|\Phi^+\rangle$ is $-1/8$, it's much less than that of $\rho_{z_0}$. As long as the Bell-like value of the measurement results are equal to $I_{\rho_{z_0}}$, Alice and Bob will not detect any insecurity in the measurement results. By using this fact, the measurement devices can fake about 99\% of the measurement results. These fake measurement results can be stored in the measurement devices without being known by Alice and Bob, and the Bell-like value of this fake part can be as less as zero. For the rest 1\%, the measurement devices choose to jointly measure (\ref{noninput}) with the built-in state $|\Phi^+\rangle$, the Bell-like value of these measurement results is $I_{|\Phi^+\rangle}=-1/8$. From the view of Alice and Bob, the average Bell-like value of the whole measurement results would be
\begin{equation}\nonumber
  I_{avaerage}=0\times 99\%+-\frac18\times1\%=-\frac1{800}
\end{equation}
which is the same as $I_{\rho_{z_0}}$. The fake part of the measurement results is known by the measurement devices, the guessing probability could be 1. For the measurement results of the built-in state $|\Phi^+\rangle$, the maximum guessing probability is $7/12$. Then the average guessing probability of the whole measurement results for the measurement devices should be $0.99\times1+0.01\times 7/12=0.9958$, and the corresponding lower bound of the output randomness is $-log_2(0.9958)=0.006$.

One thing we should note here is that, the untrusted measurement devices could not forge all the measurement results, there must be some truly random measurement results from the shared entanglement between measurement devices to make sure the average Bell-like value is less than zero. Otherwise, without sharing entanglement, the measurement devices could make $I_{\rho_{z}}<0$, which is contradict with quantum physics\cite{bra}.

According to the analysis above, the average guessing probability of the measurement results for Werner states ~\ref{werner2} should be:
\begin{equation}\nonumber
   p_{guess}=\frac{I_{|\Phi^+\rangle}-I_{\rho_{z}}}{I_{|\Phi^+\rangle}}\times 1+\frac{I_{\rho_{z}}}{I_{|\Phi^+\rangle}}\times maxP_{|\Phi^+\rangle}(a,b|\tau_s,\omega_t)
\end{equation}
The measurement correlation $maxP_{|\Phi^+\rangle}(a,b|\tau_s,\omega_t)$ may not be the optimal one, as there are some entangled states' measurement correlation larger than Bell state. We use $P_{\rho_{max}}(a,b|\tau_s,\omega_t))$ to represent the maximum correlation allowed by quantum physics, then the upper bound of the average guessing probability is
\begin{equation}
\label{pguesstrue}
\begin{split}
     p_{guess}=&\frac{I_{|\Phi^+\rangle}-I_{\rho_{z}}}{I_{|\Phi^+\rangle}}\times 1+\frac{I_{\rho_{z}}}{I_{|\Phi^+\rangle}}\times maxP_{|\Phi^+\rangle}(a,b|\tau_s,\omega_t)\\
     \leq&\frac{I_{|\Phi^+\rangle}-I_{\rho_{z}}}{I_{|\Phi^+\rangle}}+\frac{I_{\rho_{z}}}{I_{|\Phi^+\rangle}}\times P_{\rho_{max}}(a,b|\tau_s,\omega_t))
\end{split}
\end{equation}
\section{The maximum measurement correlation $P_{\rho_{max}}(a,b|\tau_s,\omega_t))$}
\setcounter{section}{2}
In this appendix, we calculate $P_{\rho_{max}}(a,b|\tau_s,\omega_t))$. It is said in \cite{chen2017} that any two-qubit density matrix $\rho_{AB}$ is an element in the linear space expanded by $\{\sigma_{i}\otimes\sigma_{j}, i,j\in{0,1,2,3}\}$($\sigma_0=\textbf{I}_{2\times2}$). $Tr(\rho_{AB})=1$, so the coefficient of $\sigma_0\otimes\sigma_0$ is $\frac{1}{4}$. Thus all the two-qubit density matrices has $15$ different coefficients, they are $c_{3^i+j}, i,j\in\{0,1,2,3\},(i,j)\neq\{0,0\}$. $\rho_{AB}$ could be written as:
\begin{equation}
  \rho_{AB}=\frac14\sigma_0\otimes\sigma_0+\sum_{\substack{i,j=0,1,2,3\\(i,j)\neq(0,0)}} c_{3^i+j}~\sigma_i\otimes\sigma_j
\end{equation}

$\rho_{AB}$ is a quantum state, which means the minimum eigenvalue $\lambda_{minAB}\geq0$. From the property of eigenvalue, the eigenvalue of $\rho_{AB}$ is less than the eigenvalue sum of its each part \cite{horn1962}:
\begin{equation}\nonumber
  \lambda_{AB}\leq\sum_{i,j=0,1,2,3} c_{3^i+j} \lambda_{3^i+j}
\end{equation}
The eigenvalues of $\sigma_0\otimes\sigma_0$ are $\{1,1,1,1\}$, the eigenvalues of $\sigma_i\otimes\sigma_j  ~(i,j)\neq(0,0)$ are $\{-1,-1,1,1\}$. We already know the coefficient of $\sigma_0\otimes\sigma_0$ is $\frac14$, then the range of $\rho_{AB}$'s eigenvalues could be represented as
\begin{equation}
  \begin{split}
     \lambda_{AB_1}\leq1/4-\sum_{\substack{i,j=0,1,2,3\\(i,j)\neq(0,0)}}~c_{3^i+j},\\
    \lambda_{AB_2}\leq1/4-\sum_{\substack{i,j=0,1,2,3\\(i,j)\neq(0,0)}}~c_{3^i+j},\\
    \lambda_{AB_3}\leq1/4+\sum_{\substack{i,j=0,1,2,3\\(i,j)\neq(0,0)}}~c_{3^i+j},\\
    \lambda_{AB_4}\leq1/4+\sum_{\substack{i,j=0,1,2,3\\(i,j)\neq(0,0)}}~c_{3^i+j}
    \end{split}
\end{equation}
Because the minimum element of $\lambda_{AB}$ is greater than or equal to zero, we have
\begin{equation}
\label{rangeofc}
  -1/4\leq\sum_{\substack{i,j=0,1,2,3\\(i,j)\neq(0,0)}}~c_{3^i+j}\leq 1/4
\end{equation}
The correlation $P(a,b|\tau_s,\omega_t)$ is the correlation sum of each basis:
\begin{equation}\nonumber
\label{combination}
  \begin{split}
  &P(a,b|\tau_s,\omega_t)=Tr[(P_a^{A'A}\otimes Q_b^{B'B})(\tau_s\otimes\rho_{AB}\otimes\omega_t)]\\
  =&Tr[(P_a^{A'A}\otimes Q_b^{B'B})(\tau_s\otimes(\sum_{i,j=0,1,2,3}c_{3^i+j}\sigma_i\otimes\sigma_j)\otimes\omega_t)]\\
  =&\frac9{16}+(\sum_{\substack{i,j=0,1,2,3\\(i,j)\neq(0,0)}}c_{3^i+j}P(\sigma_i\otimes\sigma_j))]\\
  \leq&\frac{9}{16}+\sum_{\substack{i,j=0,1,2,3\\(i,j)\neq(0,0)}}c_{3^i+j}max P(\sigma_i\otimes\sigma_j)\\
  \leq&\frac{9}{16}+\frac{max P(\sigma_i\otimes\sigma_j)}4
  \end{split}
\end{equation}
where $P(\sigma_i\otimes\sigma_j)=Tr[(P_a^{A'A}\otimes Q_b^{B'B})(\tau_s\otimes(\sigma_i\otimes\sigma_j)\otimes\omega_t)]$.

With the given POVM(\ref{povm2}) and with nonorthogonal states(\ref{noninput}), we can get $maxP(\sigma_i\otimes\sigma_j)=\frac{\sqrt{3}}{4}$, and the maximum measurement correlation is $P_{\rho_{max}}(a,b|\tau_s,\omega_t)=\frac{9+\sqrt{3}}{16}$. Then the upper bound of the average guessing probability~(\ref{pguesstrue}) is
\begin{equation}\nonumber
   p_{guess}\leq \frac{I_{\rho_{|\Phi^+\rangle}}-I_{\rho_{z}}}{I_{\rho_{|\Phi^+\rangle}}}+\frac{I_{\rho_{z}}}{I_{\rho_{|\Phi^+\rangle}}}\times \frac{9+\sqrt{3}}{16}
\end{equation}

The similar analysis could be applied to any entangled state $\rho_{AB}$ shared between Alice and Bob, thus the upper bound of the average guessing probability in our randomness generation protocol is:
\begin{equation}\nonumber
   p_{guess}\leq \frac{I_{\rho_{optimal}}-I_{\rho_{AB}}}{I_{\rho_{optimal}}}+\frac{I_{\rho_{AB}}}{I_{\rho_{optimal}}}\times \frac{9+\sqrt{3}}{16}
\end{equation}
where $I_{optimal}$ is the minimum Bell-like value for the given entanglement witness of entangled state $\rho_{AB}$.
\section{SDP output randomness}
From the perspective of the measurement devices, searching for the optimal guessing probability $p_{guess}$ is equal to solve the next optimization problem:
\begin{equation}\nonumber
  \begin{split}
    &max~p_{guess}\\
    subject~to~& \\
    p_{guess}=&\frac{I_{\rho'}-I_{\rho_{AB}}}{I_{\rho'}}+\frac{I_{\rho_{AB}}}{I_{\rho'}}\times maxP_{\rho'}(a,b|\tau_s,\omega_t)\\
    & I_{\rho_{AB}}=\sum_{s,t}\beta_{s,t}P_{\rho_{AB}}(1,1|\tau_s,\omega_t)\\
    & I_{\rho'}=\sum_{s,t}\beta_{s,t}P_{\rho'}(1,1|\tau_s,\omega_t)\\
    &I_{\rho'}\leq I_{\rho_{AB}}<0 \\
    &maxP_{\rho'}(a,b|\tau_s,\omega_t)\geq maxP_{\rho_{AB}}(a,b|\tau_s,\omega_t)
  \end{split}
\end{equation}
This problem can be formulating as semidefinite programs(SDP)\cite{sdp1996}\cite{sdp2007}\cite{c2016}, the lower bound of the output randomness obtained from SPD is shown as curve $b$ in FIG.\ref{output}.
\end{document}